\documentclass[aps,pra,showpacs,twocolumn]{revtex4}
\usepackage{graphicx}
\usepackage{bm}

\begin{document}

\title{
Nervous excitability dynamics in a multisensory syndrome \\ and its similitude with a normal state. Scaling laws}

\author{Isabel Gonzalo-Fonrodona}
\affiliation{Departamento de \'{O}ptica, Facultad de Ciencias F\'{\i}sicas, Universidad Complutense de Madrid,\\ Ciudad Universitaria s/n, 28040 Madrid, Spain}
\email{igonzalo@fis.ucm.es}

\author{Miguel A. Porras}
\affiliation{Grupo de Sistemas Complejos, ETSIME, Universidad Polit\'ecnica de Madrid, Rios Rosas 21, 28003 Madrid, Spain}
\email{miguelangel.porras@upm.es}

\begin{abstract}
In the context of increased number of works published on multisensory and
 cross-modal effects, we review a cortical multisensory syndrome (called central syndrome) associated with a unilateral parieto-occipital lesion in a rather unspecific (or multisensory) zone of the cortex.

 The patients with this syndrome suffered from  bilateral and symmetric multisensory disorders dependent on the extent of nervous mass lost and the intensity of the stimulus. They also presented cross-modal effects. A key point is the similitude of this syndrome with  a normal state, since this syndrome would be the result of a scale reduction in brain excitability. The first qualities lost when the nervous excitation diminishes are the most complex ones, following allometric laws proper of a dynamic system.

 The inverted perception (visual, tactile, auditive) in this syndrome is compared to other cases of visual inversion reported in the literature. We focus on the capability of improving perception by intensifying the stimulus or by means of another type of stimulus (cross-modal), muscular effort being one of the most efficient and least known means. This capability is greater when  nervous excitability deficit (lesion) is greater and when the primary stimulus is weaker. Thus, in a normal subject, this capability is much weaker although perceptible for functions with high excitability demand. We also review the proposed scheme of functional cortical gradients whereby the specificity of the cortex is distributed with a continuous variation leading to a brain dynamics model accounting for multisensory or cross-modal interactions.  Perception data (including cross-modal effects) in this syndrome are fitted using Stevens' power law which we relate to the allometric scaling power laws dependent on the active neural mass, which seem to be the laws governing many biological neural networks.
 \end{abstract}

\pacs{Brain dynamics, Neurophysiology, Perception, Multisensory, Cross-modal, Inverted perception, Parieto-occipital lesion, Plasticity, Facilitation, Neuropsychology, Reversal of vision, Tactile inversion, Scaling laws}

\maketitle

\section{Introduction}

In the last few years there has been an explosion of interest in
multisensory effects such as cross-modal interactions in
integrative cerebral processes (see, e.g., [1-13] and of course
the general books \cite{Libro1,Libro2,Libro3} and their
references). Multisensory interactions have been revealed by
functional magnetic resonance imaging, positron emission
tomography and analysis of blood oxygen level, which have suggested that
longstanding notions of cortical organization need to be revised.
It was determined that cross-modal interactions can affect activity in
cortical regions traditionally regarded as ``unimodal". This
occurs, for example, in the contribution of the visual cortex to
tactile perception \cite{Pascual97,Pascual04,Sathian02}. It was
then suggested that multisensory interactions must be considered
an inherent component of the functional brain organization
(e.g., \cite{Pascual04, Wallace04, Martuzzi07} and references
therein).

In this context, we shall analyze here the multisensory syndrome characterized by Gonzalo \cite{Gonzalo10, Gonzalo51,Gonzalo52}, originated
from a unilateral parieto-occipital cortical lesion equidistant
from the visual, tactile, and auditory projection areas (the middle of area 19, the
anterior part of area 18 and the most posterior of area 39, in
Brodmann terminology), as shown in Fig.1 A.
\begin{figure}
\begin{center}
\includegraphics[width=5cm]{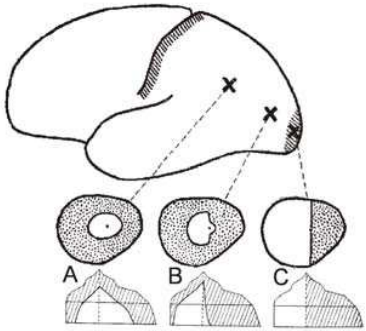}
\caption{\label{CEREBRO}Position of cortical lesions and
respective visual fields with their sensibility profile (involved zones are dark). A: central, B:
paracentral, C: marginal or peripheral, syndromes. (Modified from
Fig. (a) of \cite{Gonzalo96} with permission of the MIT Press).}
\end{center}
\end{figure}
This syndrome was called {\em central syndrome of the cortex}  to
highlight the ``central" zone of the lesion
in comparison
with the primary or projection areas (considered as ``peripheral " or ``marginal"  in
this context). The immediate repercussion of part of this research
(e.g., \cite{Critchley53,Bender48,Ajuriaguerra49}) is later addressed towards
 the development of cerebral processing models [25-32], and other directions (e.g., \cite{Arias04,Barraquer05,Sierra12}).

Among 117 selected patients with cortical lesions, 35 of them presented central syndrome of various intensities.  A comparison is made in that research \cite{Gonzalo10,Gonzalo51,Gonzalo52} between an acute central syndrome case (patient M) and a less acute one (patient T).
The syndrome is characterized by a multisensory and symmetric affection where all the sensory functions are
affected, and with symmetric bilaterality. In the visual system for example, in addition to other disorders,
there is a bilateral and symmetric concentric reduction of the visual field (Fig. 1 A), with gradation of the
involvement from the center to the periphery. The syndrome presents a functional {\em depression}, shown
for example in the excitation threshold curves for electrical stimulation of the retina
in Fig. 2 \cite{Gonzalo10,Gonzalo52},
\begin{figure}
\begin{center}
\includegraphics[width=7cm]{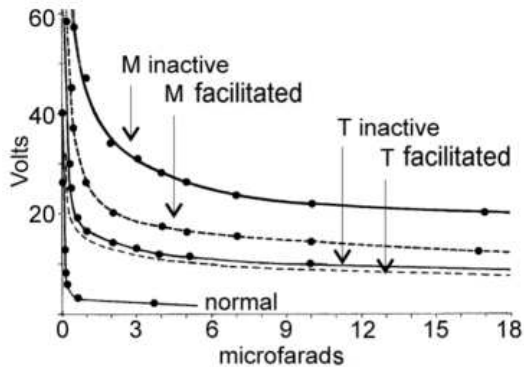}
\caption{\label{EXCITATION}Excitation threshold curves for
electrical stimulation of the retina (cathode on eyelid) in the acute
central syndrome case (patient M),  in the less intense central syndrome case (patient T), and for a normal man. For M and T cases, curves for an inactive
state (free of facilitation) and for a facilitated state by strong muscular effort. Electrical intensity (indirectly
given by volts) versus time (given by microfarads) necessary to obtain minimum luminous sensation. (From Fig. 2 of
\cite{Gonzalo09} with permission of Elsevier).}
\end{center}
\end{figure}
where the thresholds for the acute central syndrome case (patient M) are seen to be greater than those for
the less intense central syndrome case (patient T), and the latter greater than for a normal man. The functional depression
presents dynamic phenomena related to the dynamic conditions of the nervous excitability \cite{Gonzalo10,Gonzalo52}. In this chapter, we focus mainly on some aspects regarding visual and tactile systems although the involvement is general.

One of the dynamic phenomena is the functional decomposition of perception in the form of a dissociation or desynchronization of sensory qualities
normally united in perception. The sensory qualities are lost according to their excitability
demands, i.e., the higher functions (with higher excitability demand) are the first lost as the
stimulus intensity diminishes (in a given patient) or as the magnitude of the lesion grows (in several patients). Thus, the first function lost is gnosia, then acuity, blue color, etc...
In this decomposition, the direction function is manifested. For example, when the illumination of an upright white arrow diminishes, the perceived arrow is
at first upright and well-defined, next more and more rotated in the frontal plane
at the same time that of a smaller size and losing its form and colors in a well-defined physiological order \cite{Gonzalo10}.
These phenomena are treated, together with tactile and auditive inversion, in Section 2.

Among many other dissociation phenomena, the chromatic irradiation or ``flat colors" disorder (colors as separate from objects) \cite{Gelb20} was
interpreted as a dissociation between the chromatic function and the spatial localization function due to their different
excitability thresholds. In relation with orientation, the so-called {\em orthogonal} disorder, detected in patients M and T, can be considered a dissociation phenomenon in which objects are seen
the same independently of their orientation. For example, texts can be read upright or upside-down without
noting any difference. Another detected disorder in the most acute case M was the substitution of the halocentric spatial orientation by the egocentric one.

The continuous gradation found between the central syndrome and a syndrome associated to a projection path led Gonzalo \cite{Gonzalo52} to propose a functional gradient scheme in which the specificity of the cortex is distributed with a continuous variation. This model highlights a functional continuity and unity of the cortex.  The overlap of several functional specificities would account for multisensory or cross-modal interactions. Different syndromes can then be ordered according to the position and magnitude of the lesion. These ideas are the subject of Section 3.

A key point of the central syndrome is its similitude with a normal man. The sensory system maintains its organization but on a
reduced scale of nervous excitability. This fact permits us to apply the allometric laws of dynamical systems that are subjected to a scale
change, and to explain the different (allometric) loss of the different sensory functions. These points are exposed in Section 4.

Another dynamic phenomenon is the unusual capability to improve perception not only by increasing the stimulus intensity but also
by facilitation by another stimulus of the same or different modality (cross-modal effect). The added stimulus provides an
extra excitation that compensates the excitation deficit due to neural mass lost in the ``central" lesion. This capability was
found to be greater as the nervous excitability deficit (lesion) is greater and as the primary stimulus is weaker. Muscular effort
was found to be particularly efficient at improving perception (see the facilitated cases in Fig. \ref{EXCITATION}).

The available data on improvement of perception with the intensity of the facilitating stimulus, and also with the primary stimulus, are fitted using Stevens' power law, which in turn is shown to reflect the allometric scaling power laws dependent on the active mass (the biological neural network in this case). These issues are considered in detail in Sections 5 and 6.

It is also noticeable the capability of patients with central syndrome to iterative temporal summation. The slowness of the
cerebral system in this syndrome makes the cerebral excitation to a short stimulus to decay slowly. If a second stimulus arrives
before the first excitation has completely fallen down, excitations are summed up, making possible to reach the excitation
threshold to produce a sensory perception, reducing by this way the pathological dissociation.  A simple model has been proposed
recently to explain these and others temporal aspects of this syndrome, as the shortening of the perceived duration of a given
stimulus \cite{Gonzalo01,GonzaloPorras13}, and will be the subject of further work.

The rich phenomenology observed in this syndrome reveals general aspects of brain dynamics, as the functional unity of the cortex
and the continuity from lower to higher sensory functions. The syndrome and its interpretation offer a framework to understand,
at a functional level, aspects of the integrative cerebral process on a physiological basis.

\section{Inverted Perception Disorder}

The functional depression referred to in the previous section is richly illustrated by the striking inverted or tilted perception disorder in central syndrome, associated to left or right unilateral ``central" lesions \cite{Gonzalo10,Gonzalo51,Gonzalo52}.

First, we shall describe this phenomenon in the visual system.
Gonzalo \cite{Gonzalo52,Gonzalo10} reported on 25 patients with cortical lesions
that presented clearly chronic manifestation of tilted or almost inverted
vision under conditions of minimum stimulation (illumination): 13
cases between $2^{\rm o}-12^{\rm o}$, six  between $12^{\rm o}-30^{\rm o}$, and six
between $30^{\rm o}-160^{\rm o}$, all in the frontal plane, and a few cases also
with small tilt in the sagittal plane.  There were 12 with brain
injury in the parieto-occipital region, which can be considered as ``pure" central syndrome cases.

\begin{figure}
\begin{center}
\includegraphics[width=8.5cm]{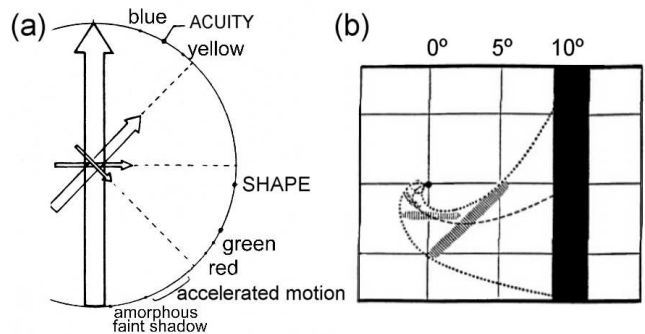}
\end{center}
\caption{\label{FLECHA}   In central syndrome: (a) Perceived vertical test arrow in the center of the visual field of right eye. Qualities loss for each inclination are indicated.
(b) Centripetal deviation of the perceived images  of a test object (black bar in the figure) situated in the right side of the visual field, at $10^{\rm o}$
from the center of the visual field. (From Fig. (a) of Ref. \cite{Gonzalo96} with permission of The MIT Press).}
\end{figure}

In the acute central syndrome of case M in an inactive state (free of sensory
facilitation), the perception of a vertical upright white arrow suffered from the following dissociation
phenomena as the cerebral excitation was diminished. During high enough illumination, the
perception of the arrow was upright, well defined and with a slight green tinge. As lighting was reduced, the arrow was
perceived as more and more rotated in the frontal plane [see Fig. \ref{FLECHA} (a)], at the same time it became smaller and with less defined shape and color, following a well-defined physiological order. The first function lost was the meaning of the object, then, blue color, visual acuity, yellow
color, red color, luminosity...  If the tilt was between $90^{\rm o}$ and $180^{\rm o}$, the object was perceived as a small shadow.
The tilt was measured by rotating the arrow in the opposite direction until it was seen upright. For a test object situated in one side of the visual field, the
object was seen to rotate with centripetal deviation, coming to rest inverted and constricted in contralateral position quite
close to the center of the visual field, as shown in Fig. \ref{FLECHA} (b). The more peripheral the vision, the more tilted the arrow was perceived to
be. In central vision, the rotation was clockwise for the right eye and counterclockwise for the left eye. In peripheral vision,
the rotation was clockwise (counterclockwise) in the right (left) side of the visual field.

The perceived turn was dependent on the size and distance of the objects, i.e., on the
subtended angle of vision, and also on the illumination and
exposure time. Thus, an object appeared to be more
tilted if it was only seen for an instant. As no inclination was
perceived in clearly distinguished (well illuminated) objects,
many patients were unaware of their anomalies, which were only
relevant when provoked in an inactive state and under low
intensity of the stimulus.

In the acute case M, the maximum inclination perceived was about
$170^{\rm o}$ with the left eye and $145^{\rm o}$ with the right
eye. In the less acute case T, the maximum inclination was only
about $25^{\rm o}$ with the right eye and $16^{\rm o}$ with the
left eye,  following the same behavior as case M. The disorder was
chronic in both cases. It was found in case T that two days after
an epileptic attack the maximum inclination was $120^{\rm o}$ with
the right eye and $70^{\rm o}$ with the left one. A slight
tendency to rotation in the sagittal plane was also detected in
case T and a few other cases. The reversal of vision was discovered
in patient M when a moving object was seen by the patient with
inverted direction of movement, and perceived as a mere blurred
spot moving along a much smaller trajectory, and with an
overestimated speed. For a moving object, the time that the object is seen in a place diminishes, and therefore the
stimulus diminishes. The inversion process was reversible, i.e., the
perception was improved by increasing the stimulus intensity or by
multisensory facilitation, as explained in Sections 5 and 6.

\begin{figure}
\begin{center}
\includegraphics[width=8.5cm]{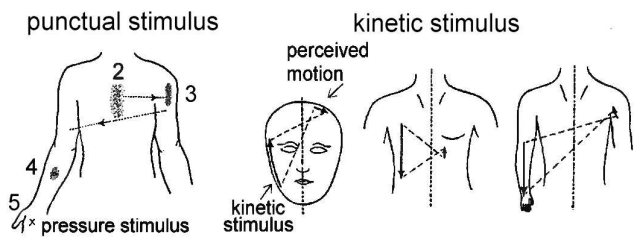}
\end{center}
\caption{\label{TACTILEINVER} In the central syndrome case M in inactive state:  Perception of a punctual stimulus on right hand. Stages or phases of tactile localization of the stimulus (see the text). Phase 3 is the inversion phase. For a cutaneous kinetic stimulus, centripetal deviation of perception towards the middle line of the body and inverted direction of motion.}
\end{figure}

Concerning the tactile system, it was found that for a mechanical pressure
stimulus on one hand, five successive stages of dissociated perception
were distinguished successively in case M as the energy of the
stimulus was increased \cite{Gonzalo10,Gonzalo51,Gonzalo52} (see Fig. \ref{TACTILEINVER} left part): 1, primitive tactile sensation without localization; 2, deviation to the middle with
irradiation (spatial diffusion in a similar way to the chromatic irradiation in vision);
3, inversion phase but closer to the middle line of the body
than the stimulus; 4, homolateral phase; 5, normal localization,
which required intense stimulus, or moderate stimulus and
facilitation by muscular effort, for example.  The lowest
sensory level stages (1, 2, 3) were separated only by very
small increments of the stimulus intensity, while the highest stages
(4, 5), close to normal localization, were separated by large
increments, these being the most perturbed stages by the cerebral
excitability deficit in the central syndrome.

When a cutaneous mobile stimulus was perceived in the inversion
phase (Fig. \ref{TACTILEINVER}), the perception was contralateral and close to
the middle line of the body, with a very shortened trajectory
(approximately 1/10 in case M) and almost inverted direction of
movement within three autonomous zones of inversion: head, upper
extremities and lower extremities.

Tactile inversion was studied for cutaneous and articular
stimulation that includes processes such as walking. In moderate
walking, the process showed striking characteristics: the first
step was ignored, the second step was felt inverted, the third
one as transversal, the fourth oblique, etc.  This is a
progressive recruitment in the direction of the perceived steps due
to accumulation of excitation in the nervous centers by iterative
action of steps due to the capability of iterative temporal summation. In slow walking, there was no summation and the direction remained inverted for
each step, the steps being felt as very short.

It was found that the general laws for the tactile direction were the same as for
the visual direction function. This led Gonzalo \cite{Gonzalo10,Gonzalo52} to consider spatial
inversion in a general way and as an essential fact in the
organization of sensory functions with a spatial character, and to postulate a continuity between lower and higher sensory functions. Together with the turn, there is a diminution of sensorial {\em intensity, space} and {\em time}, dynamic parameters whose values depend on the active neural mass \cite{Gonzalo10,Gonzalo51,Gonzalo52}.

With respect to the auditive system, cases M and T presented also
a dissociation phenomenon in analogous
way as in vision and touch. For a particular sound, they perceived
simple sonorousness if the stimulus is weak enough, and real tone if the stimulus is more intense.
Contralateral localization due to spatial inversion of a sound
stimulus occurred only in case M when the stimulus was weak and
the patient was in an inactive state. The inverted perception always
lacked tonal quality \cite{Gonzalo52}.

Other cases were reported by Gonzalo \cite{Gonzalo10,Gonzalo52}
that presented a sudden and transitory tilt or reversal of vision,
mainly during epileptic auras. In this type of sudden turn, the
visual scene was not as deteriorated as in the earlier case. He
observed different degrees of tilted vision not only in cases with
lesions on parieto-occipital region but on the occipital pole, and
also far from the occipital area (e.g., a very anterior
parieto-temporal region), showing that the anomaly occurs not
only in central syndrome. He stated \cite{Gonzalo10} that the
disorder in the visual direction function is not an autonomous
syndrome, but it is connected with the rest of visual functions;
and could be present in cases with cerebral lesions in different
locations, provided there exists some involvement of visual
functions. Moreover, this disorder could be a rather common
affection since the direction function is easily perturbed in
different types of lesions, giving place, at least, to small
inclinations of the visual image.  Tilts in chronic disorder can
be only evidenced under convenient exploration since a good
illumination of the objects and sensory facilitation make the
disorder go unnoticed for the patient very often. In the
bibliography revised by Gonzalo up to 1950 he found about 27 cases
with permanent disorder of tilted vision (in the sense as already described),
and 45 cases which presented transitory turns during
attacks. In general, in the bibliographic revisions made until
recent years \cite{Solms88,Gonzalo07,Sierra12} most of the cases
are transitory, without loss of shape and size of the perceived
object or the visual scene, except in very few cases (e.g., some
similar features in the degradation of the image are reported in
\cite{Kasten06}). Both permanent and transitory types are
associated to a wide variety of cerebral lesions but with
predominance of the parieto-occipital or
parieto-occipital-temporal regions, and only very few cases (two or three cases) are
associated to the frontal region. There are also some cerebellar
cases, apart from other etiologies such as multiple sclerosis and
epilepsy.

Concerning the rotation plane, in almost all cases reported in the
literature and in those described by Gonzalo, the rotation of the
visual image was in the frontal plane. In one of the patients of
River et al. \cite{River98} the sense of the rotation of the
visual scene was specified to be counterclockwise for the left
visual field, as described before. In general, few cases (e.g.,
[42-45]) presented rotation in the sagittal plane, and some of
them presented in addition rotation in the frontal plane, as in
case T. In a few other cases [e.g., \cite{Gertsmann26}
(one case), \cite{Arias01} (four cases)], the image was rotated in
the horizontal plane, leading to left-right visual inversion.

As for the orthogonal disorder in the central syndrome, in which
objects upright and upside-down seem the same without noting any
difference, a similar phenomenon was reported  by  Solms et al.
\cite{Solms88}.

To the best of our knowledge, tactile and auditive inversions as
described here (with similar laws to those of visual inversion)
have not yet been reported by other authors. This type of tactile
inversion must be distinguished from the frequently described
tactile allochiria (tactile allesthesia). In the latter,
analogously to visual allochiria or allesthesia, tactile
localization is contralateral to the tactile stimulus, in general
symmetric (without centripetal deviation), and from the healthy
side to the affected side (\cite{Marcel04,Kawamura87} for
example).

\section{Cortical Functional Gradients}

\begin{figure}[!t]
\begin{center}
\includegraphics[width=7.5cm]{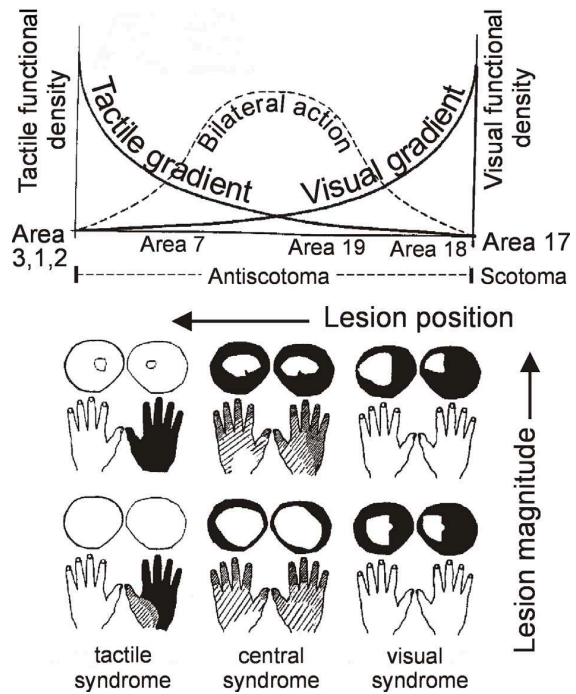}
\end{center}
\caption{\label{GRADIENTES} Lower part: Visual fields and tactile
sensitivity of 6 cases studied by Gonzalo \cite{Gonzalo10}, ordered according to the position and
magnitude of the lesion. The degree of the defect is greater in
darker regions. Upper part: scheme of visual and tactile
functional densities, and the unspecific functional gradation
which represents the multisensory and bilateral effect. Brodmann
areas are indicated. (Adapted from Fig. 5 of \cite{Gonzalo96} with
permission of The MIT Press).}
\end{figure}

The gradation found between the central syndrome and a syndrome of
the projection paths; and between different central syndrome
cases of various degrees, led Gonzalo to define two types of
continuous functions through the cortex \cite{Gonzalo52,Gonzalo09}, shown in Fig. \ref{GRADIENTES}. One
type describes the specific sensory functional densities, of
contralateral character, with a maximum value in the respective
projection area and decreasing gradually towards a more ``central"
zone and beyond so that the final decline must reach other specific areas,
including their primary zones. This is illustrated in Fig. \ref{GRADIENTES}, where
the specific visual function density
reaches all the tactile area until its primary zone.
 This type of function combines the
factors of position and magnitude of a lesion.
 The more ``central" the lesion, the greater the lesion must be
to originate a specific anomaly of the same intensity.
For a given position of the lesion, its magnitude
determines the degree of functional depression. The other type of
function has an unspecific (or multispecific) character, is maximum
in the ``central"   region (where the decline of the earlier
mentioned specific functions overlap) and vanishes towards the
projection areas. It represents the multisensory effect in the
anomalies and the bilaterality or interhemispheric effect by the
action of the corpus callosum. Each point of the cortex is then
characterized by a combination of specific contralateral action
with unspecific ``central"  and bilateral action.

The central syndrome refers, as stated earlier, to lesions in the ``central" zone,
equidistant from the visual, tactile, and auditive projection paths. This syndrome exhibits both maximum multisensory and dynamic effects. The concentric reduction of the visual field is called  anti-scotoma by contrast to the scotoma cases, as indicated in Fig. \ref{GRADIENTES}.
Two central syndrome cases of different magnitude are shown in the lower part of this figure.
Syndromes corresponding to lesions in the projection paths
suffer from a functional suppression restricted to
the contralateral half of the corresponding sensory system and
scarcely present dynamic effects. For example, in the visual system there is a loss of the contralateral visual field, as was shown schematically in Fig. 1 C. Intermediate syndromes between these and central syndromes were called ``paracentral" syndromes, bilateral involvement being asymmetric. In a visual paracentral syndrome, for example, there is an asymmetric concentric reduction of the visual field.
Concerning the two tactile syndrome cases shown in the left part of the Fig. \ref{GRADIENTES}, only a very slight visual defect is present in the case where the magnitude of the lesion (in the tactile region in these cases) is high enough.

\begin{figure}[!ht]
\begin{center}
\includegraphics[width=6cm]{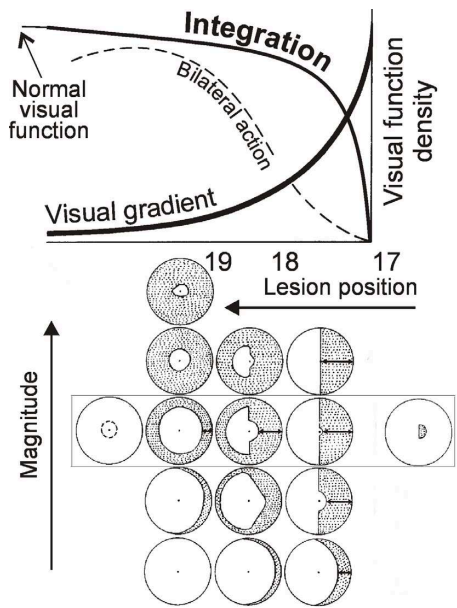}
\end{center}
\caption{\label{GRADVIS} Scheme of the visual gradient and its integration. Visual field patterns ordered by position and magnitude of the lesion, the narrowing of the affected field being indicated. A series is highlighted in a box to stress the successive type of syndromes. Concentric reductions correspond to central syndromes.}
\end{figure}

Fig. \ref{GRADVIS} shows more specifically the effect of the position and magnitude of the lesion in the visual field, and the corresponding visual gradient. It must be understood that, for the visual function to be normal, the action of the region with greatest visual sensory function density is not enough, and the whole specific functional
density in gradation through the cortex must be involved in the integrative cerebral process, leading to the normal sensory visual function, as shown schematically in Fig. \ref{GRADVIS} by means of the integration curve.  It is noticeable, for example, the significant participation of the traditionally  ``extravisual"   cortex in the maintenance of the visual field. The same applies to other senses and qualities. In a formal and abstract way, it can be said that the specific functional density is related to the derivative of the integration through the cortex (upper curve in Fig. \ref{GRADVIS}), which justifies that the specific function was named
gradient. The specific gradient would take into account the density of specific neurons through the cortex and their connections,
representing the dynamic aspect of its anatomic basis. A sensory signal in a projection area would be only an inverted and constricted
outline that must be magnified and reinverted, i.e., integrated over the whole region of the cortex where the
corresponding specific sensory functional density is extended. Magnification would be due to the increase in recruited cerebral
mass, and reinversion due to some effect of cerebral plasticity, following a spiral growth, as in Fig.
\ref{FLECHA} \cite{Gonzalo10,Gonzalo51,Gonzalo52}. In the visual system, reinversion and
bilateralization would occur in the 18 and 19 Brodmann areas, where the sensory representation is already reinverted.
We must note that the primary or projection zones, where the respective specific functional
gradients are maximum, are highly organized and specialized while the ``central" zone  is rather unspecific with
capacity for adaptation or learning. It is very small in animals, and even in other mammals, but it has a large extension in man.

In this framework, the central syndrome was interpreted as a deficit of cerebral integration due to a deficit of cerebral nervous excitation caused by the loss of a rather unspecific (or multi-specific) neural mass \cite{Gonzalo10}. Thus, it can be considered as the result of a deficitary nervous excitation of the cerebral system.

To conclude this section, we can say that, contrary to the rigid separation of regions (mosaic type), the functional gradients account for a functional continuity and physiological
heterogeneity of the cortex, this one being subjected to a common principle of organization. The gradients scheme is an abstraction of the
observed facts and offers a dynamic conception of {\em quantitative} localizations which permits an ordering and an interpretation of
multiple phenomena and syndromes. A very similar gradients scheme was proposed by Goldberg \cite{Goldberg89}. The model described here is in close connection with the gradients found in the last years by means of neuroimage techniques \cite{Tal09,Hertz10,Amedi05}, and with findings and proposals based on a distributed character of the cerebral processing and its adaptive aspects [54-59, 18].
It was suggested that traditional specific cortical domains are separated from one another by transitional multisensory zones and that multisensory
interactions occur even in the primary sensory cortices
\cite{Martuzzi07,Kayser07,Sathian02}. The functional gradients model also accounts for multisensory interactions, which is a requirement formulated by several authors.

\section{Similitude and Allometric Scaling Power Laws}

An essential feature of the central syndrome that makes it of special interest is its similitude with a normal case, since, as we show in the following, the syndrome can be interpreted as the result of a scale reduction in the cerebral nervous excitability with respect to a normal  cerebral system.

\begin{figure}
\begin{center}
\includegraphics[width=6cm]{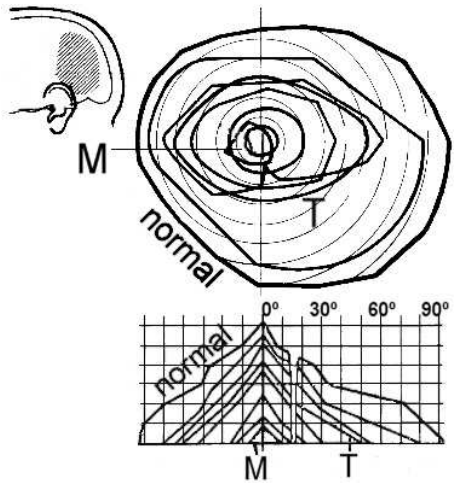}
\end{center}
\caption{\label{SENSIBILITY} Concentric reduction of visual fields and their sensibility profiles, for several central syndrome cases (M and T indicated) and a normal man. Lesion zone is indicated upper left. }
\end{figure}

The functional depression originated in the central syndrome is the result of a new dynamic equilibrium which
maintains, nevertheless, the same cerebral organization as in a normal case.
This can be appreciated, for example, in the hypoexcitability of the nervous centers
in the excitability curves of Fig. \ref{EXCITATION} which, having the same shape, are shifted with respect to the normal case following the same law. The same occurs for the luminosity
threshold curves \cite{Gonzalo10,Gonzalo52}, and also for the concentric reduction of the
visual fields (anti-scotoma) and their sensibility profiles, that maintain approximately the same shape as in a normal case but in a reduced size, as seen in Fig. \ref{SENSIBILITY}. The same can be said for the visual acuity profiles and for the rest of sensory functions or qualities. In fact, the same type of electrical excitability threshold curves as those shown in Fig. \ref{EXCITATION} were found for the tactile system \cite{Gonzalo10,Gonzalo52}. The cases of central syndrome with different degrees due to the different magnitude of the lesions, are then not only similar to a normal case but they are also similar between each case (see Fig. \ref{SENSIBILITY}), being related between themselves by a scale change.

The concept of dynamic similitude, according to which different parts of a dynamical system change differently under a change in the size of the system, was then applied to the central syndrome \cite{Gonzalo10}.  It is known that in the growth of a biological system, the sizes of two parts (say $x$ and $y$) of the system are approximately related by a scaling power law of the type
\begin{equation}
 y = A x^n ,     \label{POWER}
 \end{equation}
$n$ being different for different parts ($y_1, y_2$, ...) of the
system. These parts change then differently, i.e., allometrically
\cite{Perkkio85}. This power relation means that the rates of
growth of the two parts compared are proportional, i.e.,
$(1/y)(dy/dt)=n(1/x)(dx/dt) $ [as obtained from Eq. (\ref{POWER}) by
taking the logarithm and differentiating]. These ideas were
applied to the sensorial growth (or reduction), and an allometric
variation of the different sensory qualities was then proposed.

\begin{figure*}[!t]
\begin{center}
\includegraphics[width=13cm]{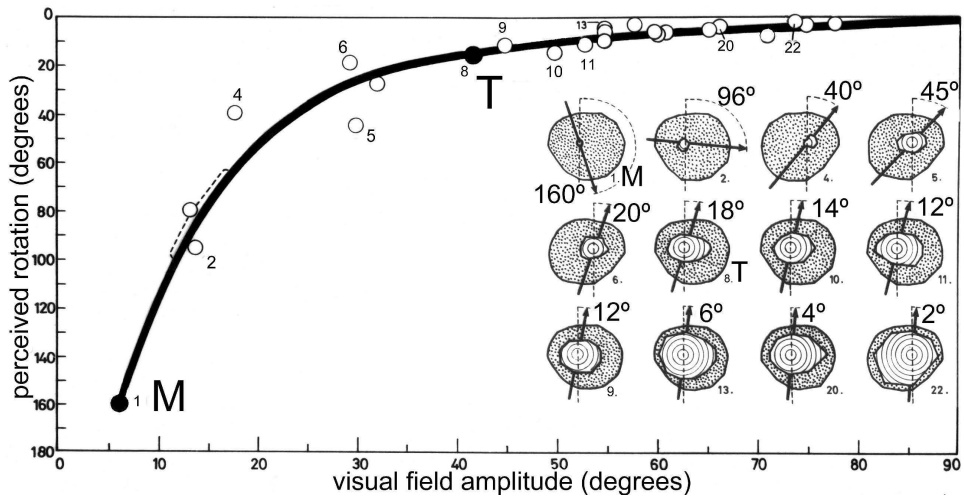}
\end{center}
\caption{\label{GIROS} Correlation between the perceived rotation of a vertical upright test arrow and the amplitude of the visual field, for 24 central (M and T are indicated) and paracentral (asymmetric) syndrome cases with permanent tilted perception disorder. On the right, the visual fields of the 12 pure syndrome cases are indicated with a small number, and the arrows indicate the rotation degree. Although the rotation sense of the arrow is opposite for left and right eyes, the same sense of rotation is drawn in the figure for a better comparison of the absolute values of rotation between the different cases. (Adapted from Fig. 17 of Suplemento II in Ref. \cite{Gonzalo10} with permission of the Red Tem\'atica en Tecnolog\'{\i}as de Computaci\'on  Artificial/Natural and the University of Santiago de Compostela, Spain).  }
\end{figure*}

In this regard, it is very remarkable the correlation found between the tilt of the perceived image (it could be variable $y$) and the width of the visual field (it could be variable $x$) in 24 cases studied by Gonzalo, with permanent tilted perception disorder, as shown in Fig. \ref{GIROS} \cite{Gonzalo10,Gonzalo52}. Among these 24 cases, 12 were considered as ``pure" central syndromes of different intensity (those shown in the right part of Fig. \ref{GIROS}), and then a power allometric law of the type  (\ref{POWER}) is expected.

\begin{figure}[!ht]
\begin{center}
\includegraphics[width=6cm]{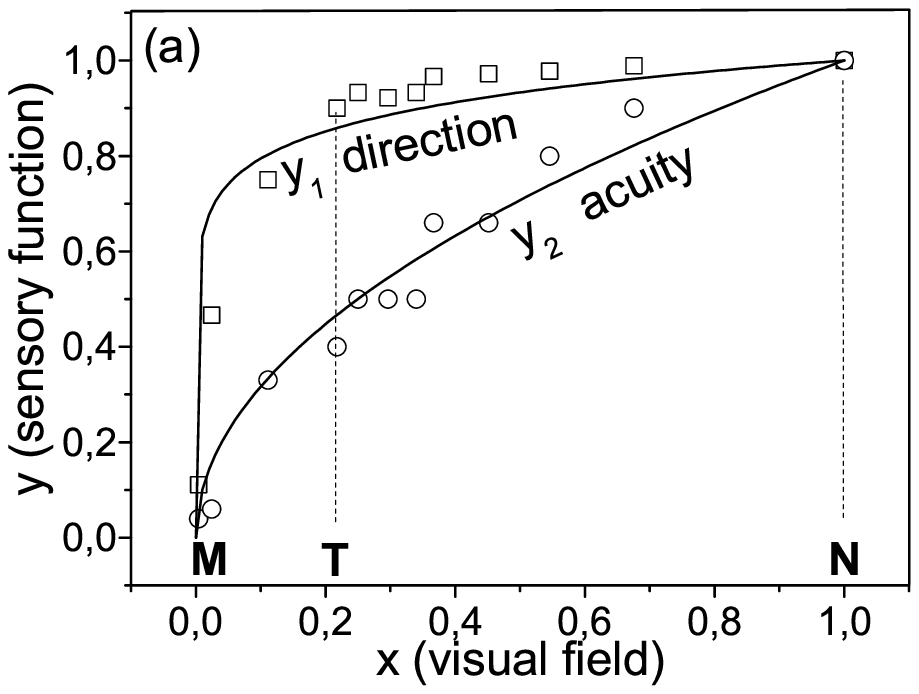}
\hspace*{0.3cm}
\includegraphics[width=6cm]{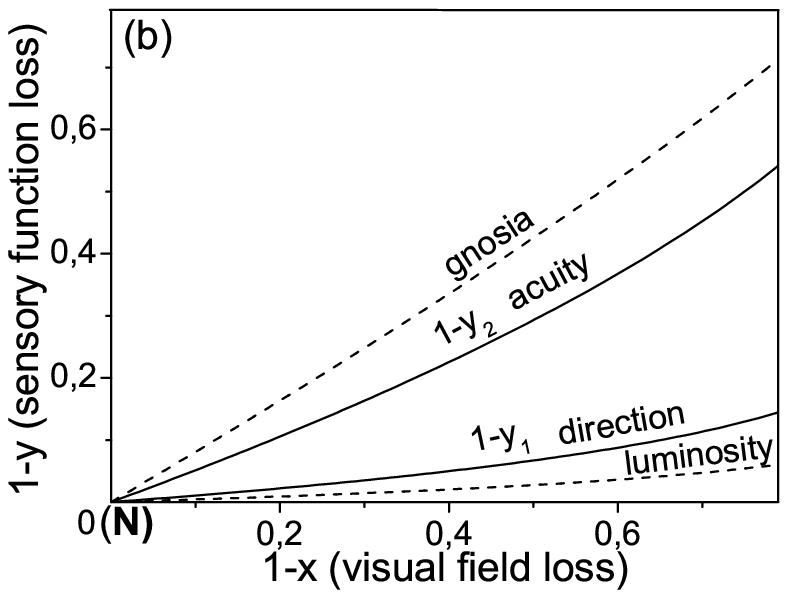}
\end{center}
\caption{\label{ALLOMETRIC} For central vision: (a) Normalized direction function
(squares) and normalized visual acuity (circles)
versus normalized untouched visual field surface $x$ of the
observing eye. Respective fittings (solid curves) with $y_1= x^{0.5}$  and  $y_2= x^{0.1}$. Cases M, T and normal (N) are indicated. (b) Curves
for the loss of direction function, $1-y_1$, and loss of visual acuity,
$1-y_2$, versus loss of the visual field, $1-x$, for the same
conditions and normalizations as in (a). Qualitative indications
for the loss of the higher sensory function gnosia and lower
sensory function luminosity are shown as examples. (From Fig. 7 (a) and (b) of Ref. \cite{Gonzalo09} with permission of Elsevier).}
\end{figure}

Indeed, we show in Fig. \ref{ALLOMETRIC} (a) the allometric power laws of type (\ref{POWER}) that fit to the data for these 12 cases. They relate the visual direction function (denoted by $y_1$) to the healthy visual field surface (denoted by $x$), and the acuity function (denoted by $y_2$) to the visual field $x$ \cite{Gonzalo09}.
In Fig. \ref{ALLOMETRIC}, the direction function is considered $0^{\rm o}$ for the total
inverted perception of the upright test arrow, and $180^{\rm o}$ for the
upright perception (normal), i.e., the maximum value is achieved
in a normal integrative process from the inverted signal in the
projection path. We consider normalized values with respect to the normal value, i.e.,
for a normal case N the normalized values $x$,
$y_1$ and $y_2$  are the unity, the maximum value. For case M, the
corresponding values are very small since the non-normalized
values are, 6 degrees for the visual field width, 0.04 for the
acuity and $160^{\rm o}$ for the perceived direction.
These normalized values of $y_1$,  $y_2$ and $x$ are used in Fig. \ref{ALLOMETRIC} (a). The acute
case M (with considerable neural mass lost), the less intense case
T (with less neural mass lost than M) and a normal case N are
indicated. Errors are greater in cases with intense central
syndrome because of their high sensory degradation and the high
capability to facilitation by temporal and multisensory summation.
Since $y_1$ and $y_2$ are 1 for
$x=1$, the value $A$ of Eq. (\ref{POWER}) is 1 and the approximate
allometric power laws we found are $y_1= x^{0.5}$ for the direction
function and $y_2= x^{0.1}$ for the acuity function. If the widths
of the visual fields are considered instead of the surface values
$x$, there is no evidence of power laws.

In order to better appreciate the allometric loss of the sensory functions in the
pathological disgregation of the sensorium in central syndrome,
we show the loss of  direction function $(1-y_1)$ and the loss of visual acuity $(1-y_2)$ versus
the loss of visual field $(1-x)$ in Fig. \ref{ALLOMETRIC} (b).
A qualitative representation of the loss of an elementary function such as luminosity, and a higher one as gnosia, according to the observed facts, is included. The origin of the graphs corresponds to a normal man (N). We can see that for a particular loss of visual field (due to a given central lesion), a split of the different qualities occurs so that the
higher  ones (e.g., gnosia) are loss in a higher degree than the lower (elementary) ones (e.g., luminosity). The order of the splitting
corresponds to the order of complexity (directly related to the excitability demand) of the sensory functions and then to the order they are lost due to the
shifts in their threshold excitabilities.

The most complex qualities, with the greatest nervous excitability (and integration) demand, become lost or delayed in
greater degree than the most simple ones (with lower excitability
demand). Sensations usually considered as elementary are then seen
to be decomposed into several functions, one of them being the
direction function, thus revealing up to a certain extent the
organization of the sensorium. Very small differences in the
excitation of different qualities already occur in the normal
individual (in colors for example), and they grow considerably in
central syndrome, as the allometric laws show. In normal individuals, there is
already a significant dependence of the visual acuity with the luminous intensity, that is much more pronounced in a patient with central syndrome. In this context, it could be said that the
cerebral system of the normal man works like an almost saturated
cerebral system, in the sense that a very low stimulus induces cerebral
excitation enough to perceive not only the simplest sensory
functions but also the most complex ones in a synchronized way.

Concerning the tactile system for a central syndrome, the tactile qualities with
the greatest demand of nervous excitation were the first lost as the
intensity of the stimulus was diminished. For example, the first quality
lost was temperature, then pain, and then pressure.

\begin{figure}[!ht]
\begin{center}
\includegraphics[width=7cm]{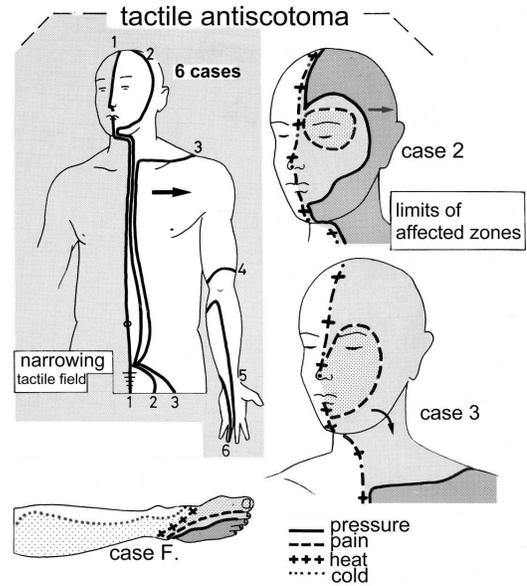}
\end{center}
\caption{\label{1TACTILE} Narrowing of the tactile field in 6 cases of tactile paracentral syndrome with different intensity. Solid line: limit to perceive a given pressure, from the most acute case 1 (very narrow tactile field) to the less acute case 6. Different limits for the loss of different tactile qualities (cold, heat, pain, pressure) for case 2, case 3, and case F (allometric dissociation).  Darker parts are more affected (anti-scotoma). Similitude between the respective limits of pressure and pain in cases 2 and 3.   (Adapted from Fig. 24 of Suplemento II in Ref. \cite{Gonzalo10} with permission of the Red Tem\'atica en Tecnolog\'{\i}as de Computaci\'on  Artificial/Natural and the University of Santiago de Compostela, Spain).}
\end{figure}

\begin{figure}[!ht]
\begin{center}
\includegraphics[width=7cm]{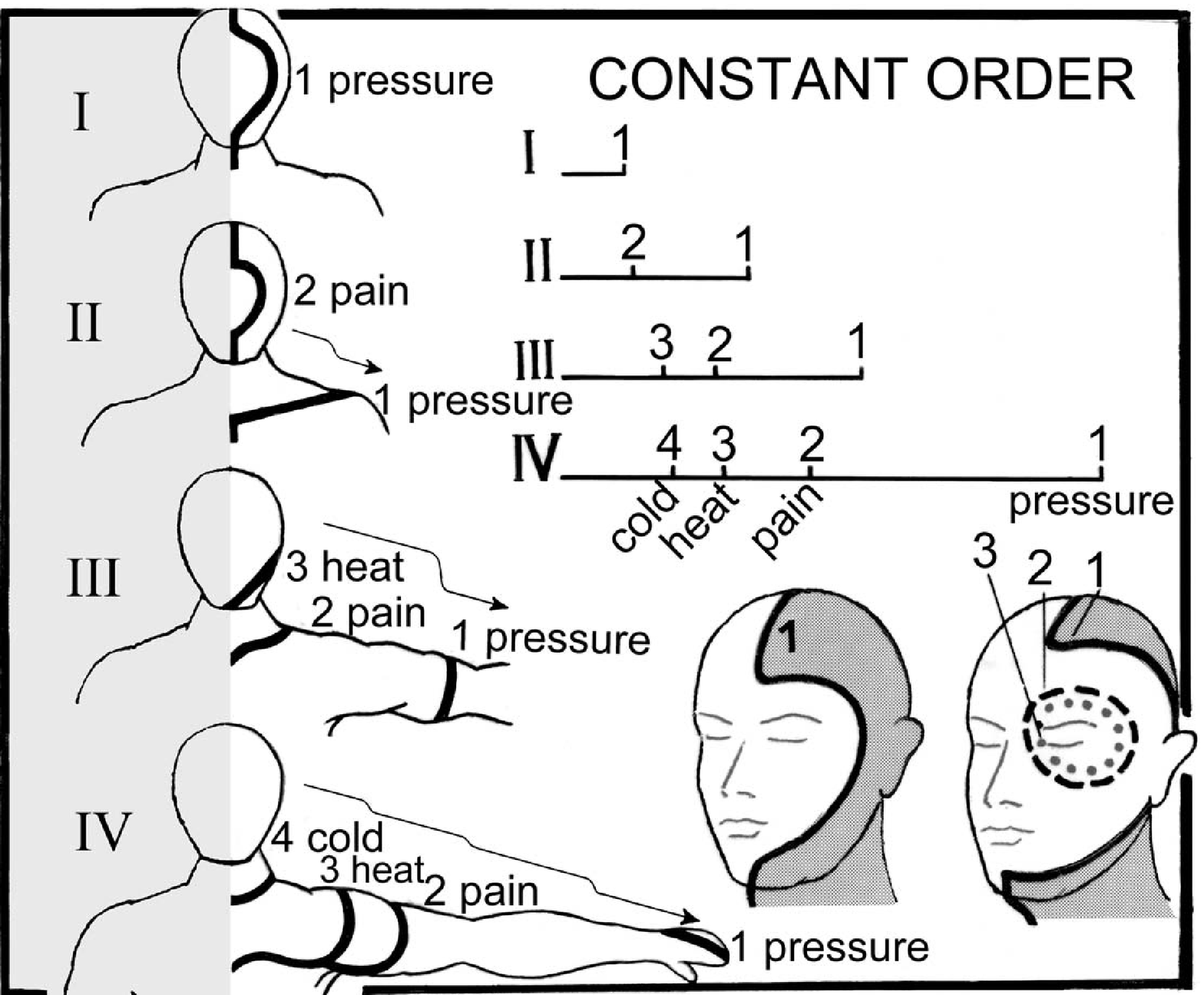}
\end{center}
\caption{\label{2TACTILE} Constant order in the loss of qualities: cold (4), heat (3), pain (2) and pressure (1), for the tactile paracentral cases I, II, III and IV (decreasing affection from I to IV).  Limits for the loss of the qualities are indicated. The most intense affection is towards the periphery (indicated by arrows).   The progressive distance between the limits is shown in horizontal lines for these four cases. The same order is shown in facial zone for two different cases. (Adapted from Fig. 25 of Supplement II in Ref. \cite{Gonzalo10} with permission of the Red Tem\'atica en Tecnolog\'{\i}as de Computaci\'on  Artificial/Natural and the University of Santiago de Compostela, Spain).}
\end{figure}

To illustrate the allometric behavior in the tactile system,
Fig. \ref{1TACTILE} shows the narrowing of the tactile field in six cases studied by Gonzalo \cite{Gonzalo10, Gonzalo52}, corresponding to different degrees of tactile paracentral cases (i.e., asymmetric central syndrome with predominance of tactile involvement in the contralateral side to the lesion). Analogously to the visual field, the so-called {\em tactile anti-scotoma} accounts for the progressive growth of the affection towards the periphery, a tendency to bilaterality, and shows dynamic effects.
In  Fig. \ref{1TACTILE}, the most acute case is 1, and the less acute case
 is 6. The solid line corresponds to the limit of the perception of the most elementary quality: tactile pressure. In cases 2 and 3 (case 2 more acute than 3), the limits of perception of the different qualities are shown. The similitude of the shape of the limits (pressure and pain in these cases) is remarkable.  Each quality is affected in a different degree according to its excitability demand, leading to the allometric dissociation between different qualities (see also the extremity of case F. in Fig. \ref{1TACTILE}).
As can be appreciated in the tactile paracentral cases I - IV shown in Fig. \ref{2TACTILE}, the different qualities are lost in a constant order. The involvement decreases from case I to  IV, the most intense affection being towards the periphery, indicated by arrows \cite{Gonzalo10}.  The sensation of cold is the first quality lost, i.e., the slightest disorder involves loosing cold sensation only, while the loss of pressure sensation implies the loss of other higher tactile qualities.  The progressive distance between the limits of the lost qualities is shown in horizontal lines. The same order and relative distances are shown in the facial cases in the lower right of Fig. \ref{2TACTILE}.

The maintenance of the shape of the different boundaries for the
loss of different tactile qualities is analogous to the maintenance of the shape of the isopters in the visual fields of central syndromes, i.e., there is an isomorphism between involvements at different degrees, governed by the allometric laws.
In summary, the allometric behavior (different exponent for each quality)
accounts for the formal and quantitative description for the loss of qualities in the central syndrome, as result of its similitude to a normal cerebral system but on a reduced scale in the nervous excitability.

\section{Improving Perception by Means of Intersensory or Cross-modal Facilitation. Power Laws}

Next, we analyze the dynamic effect which consists in improving perception through the facilitation of multisensory interactions such as cross-modal effects.

Cross-modal effects were reviewed earlier by Hartmann
\cite{Hartmann35}in a
fundamental book where, among several references to different types of multisensory
interactions, reference was made to the improvement of visual acuity in
subjects with damaged brain \cite{Urbantschitsch88} and in normals \cite{Kravkov30,Hartmann33} by auditive stimuli,
as well as by other stimuli \cite{Hartmann33}.
It is also remarkable the amount
of work in the early thirties on intersensory relations in Soviet
Union (reviewed by London \cite{London54}), and the extensive study of Gonzalo on multisensory interactions including
cross-modal effects in visual, tactile, or auditive perception in subjects with lesions
in the parieto-occipital cortex \cite{Gonzalo10,Gonzalo51,Gonzalo52}.
In the last decade, cross-modal effects and multisensory
integration has become a highly and increasing active research
topic.
For a rather recent review of cross-modal influences of sound and touch
on visual perception, see, for example, Ref. \cite{Shams10} and references therein. More
general reviews are in Refs. \cite{Libro1,Libro2,Libro3}.

By contrast, there are very few studies on the effects of motor system on
perception  except the research by
Gonzalo \cite{Gonzalo10,Gonzalo51,Gonzalo52}. This author performed a wide and quantitative study on improvement of
perception by muscular effort, as well as by other stimuli in other sensory modalities, in patients with central syndrome. In these patients, disorders such as tilted or inverted perception, concentric
reduction of the visual field, loss of visual acuity, for example, were noticeably reduced by muscular effort or
by means of other stimuli. This author also gave an interpretation to the injured brain {\em Schn.} case \cite{Goldstein18} who was able to recognize objects thanks to head movements.

\begin{figure}
\begin{center}
\includegraphics[width=6.5cm]{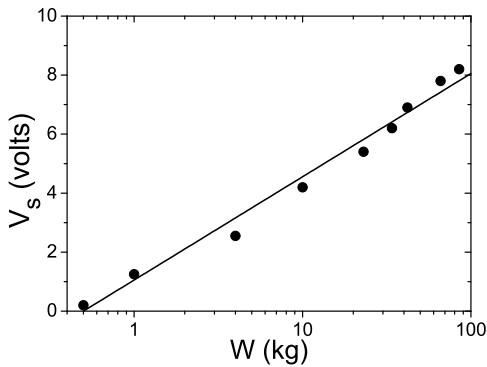}
\caption{\label{REINFORCEMENT} Saved voltage $V_s$ (volts) in electrical stimulation of
retina versus static muscular effort made by holding weights $W$ (kg in log scale), to obtain minimum phosphene, for the acute central syndrome case M \cite{Gonzalo10}. (From Fig.1 of Ref. \cite{Neurocomputing13} with permission of Elsevier).}
\end{center}
\end{figure}

We note in Fig. \ref{EXCITATION} that the curve for M in the facilitated state by muscular effort
descends towards the curve for a normal man. In relation with this experience,
it is of particular interest the quantitative equivalence found between stimuli as different
as muscular effort and electrical excitation of retina, to produce minimum
phosphene \cite{Gonzalo10}, as shown in Fig. \ref{REINFORCEMENT}  \cite{Neurocomputing13}.
Each point in this figure represents the saved voltage $V_s$ in electrical stimulation of retina to obtain minimum phosphene;
e.g., the higher point in this curve (8 volts saved) means that 12 volts instead of
20 volts were needed when the patient M was holding a weight $W = 80$ kg (40 in each
hand). The logarithmic dependence $V_s \propto \log W$ means that
a given increment in weight is much more efficient at saving voltage for smaller weights.

It was determined that muscular effort was one of the most efficient ways to improve the perception for any of the sensory
systems. This study revealed that the greater the muscular innervation involved the greater facilitation was obtained. A remarkable general fact was that the improvement by cross-modal effects was greater as the primary stimulus to be perceived was weaker, and as the cerebral lesion was greater, i.e., as the deficit in the cerebral
excitation was greater \cite{Gonzalo10}. Thus, the conclusion for a normal individual was that
facilitation cross-modal effects should be much more weaker or even negligible, and difficult to put in evidence.

\begin{figure*}
\begin{center}
\includegraphics[height=4.1cm]{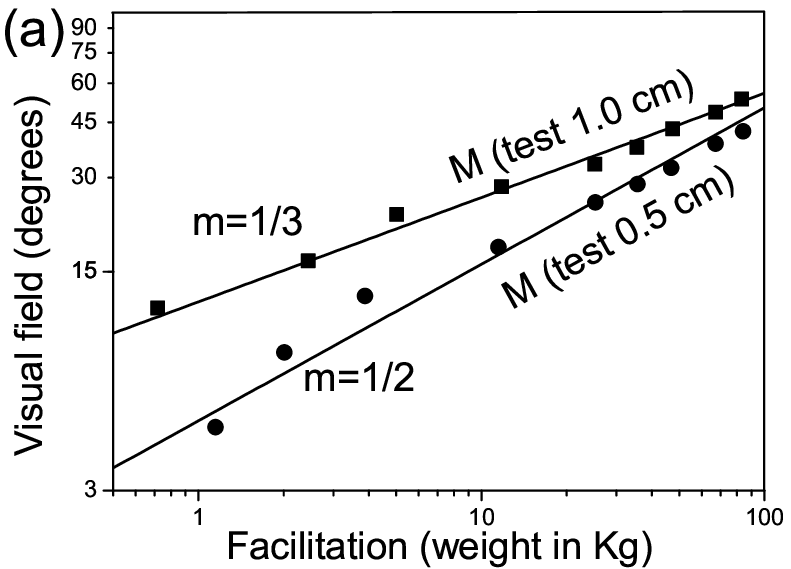}
%\vspace*{0.2cm}
\includegraphics[height=4.1cm]{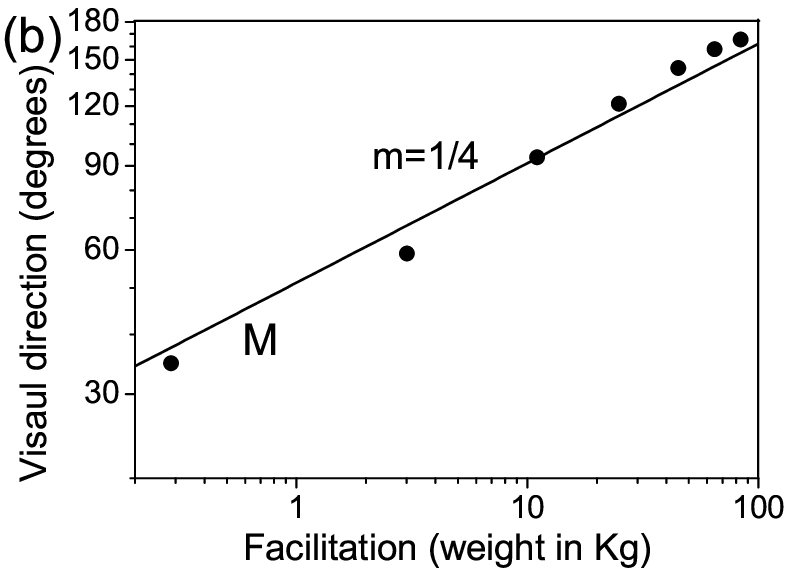}
%\hspace*{0.2cm}
\includegraphics[height=4.1cm]{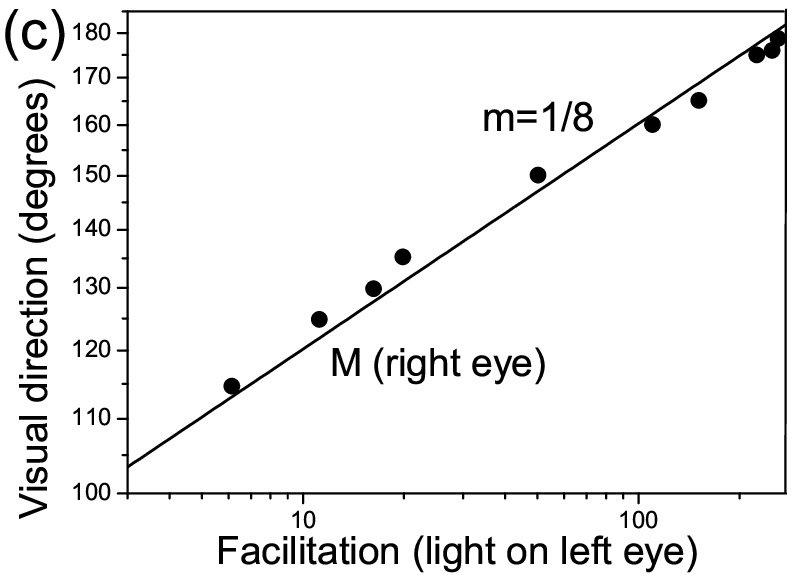}
\end{center}
\caption{\label{PLAWFACIL} For right eye of case M, log--log
representation of (a) visual field amplitude versus facilitation by
muscular effort holding weight in his hands [squares: 0.5 cm
diameter test size, dots: 1.0 cm diameter
test size], (b) visual direction (reinversion)
versus facilitation by muscular effort, (c)
visual direction versus facilitation by illumination on left eye. Exponents $m$ are indicated.}
\end{figure*}

By using available data from the patient M \cite{Gonzalo10}, we show in Fig. \ref{PLAWFACIL}
some examples of improving perception versus the intensity of a facilitating stimulus, whereas
the intensity of the primary stimulus was maintained at a constant low value.
All data correspond to a stationary regime (the facilitating stimulus was acting until a stationary perception was reached).
The well-known Stevens' power law \cite{Stevens57} that establishes a relationship between  stimulus, $S$, and  perception $P$,
\begin{equation}
P= p\,S^m  \label{STEVENS}
\end{equation}
(considered an improvement to the Fechner law) was used for fitting the
data \cite{Gonzalo09}, the slope of the straight line in a log-log representation
being the exponent $m$.
Fig. \ref{PLAWFACIL} (a) shows, in log-log representation, the
visual field amplitude of the right eye of patient M, versus facilitation
by muscular effort when holding in his hands increasing weights \cite{Gonzalo10,Gonzalo52}.  The
data fit  to straight lines with slopes  $1/2$ and $1/3$ for the
respective  diameters, 1 cm and 0.5 cm, of the white circular
test object. A greater test object implies greater stimulus, and
we can see that it leads to a lower slope, i.e., to a lower effect
of the facilitation according to what was said above. Figure \ref{PLAWFACIL} (b) shows data under
similar conditions as in (a) but the sensory
function measured is the recovery of the upright direction
($180^{\rm o}$) of an upright test white arrow that the patient perceived
tilted or almost inverted ($0^{\rm o}$º) under low illumination. The data
fit to a power law of exponent $m = 1/4$. The novelty in Fig. \ref{PLAWFACIL} (c) is that
facilitation is supplied by illuminating the left eye which is
not observing the object, the power obtained from the fitting being  $m=1/8$
\cite{Gonzalo09, Gonzalo07P}.

There are also some cases reported in the literature \cite{Gonzalo07} in which certain
stimuli as eye closure, changes in body position, moving
one hand, seeing a flame or holding on to a fix object, corrected
and reinverted the vision \cite{Solms88}(one case), \cite{Penta49} (one case), \cite{Klopp51} (three cases), \cite{Potz43} (one case),
\cite{Arjona02} (one case), \cite{Cohen91} (one case), \cite{River98} (one case).
Facilitation  cross-modal effects have also been observed in patients with
other visual deficits, e.g., \cite{Frassinetti05,Poggel06}, and also in
normals (e.g.,  \cite{Martuzzi07,Gillmeister07,Diederich07,Rowland07,Baier06,Schnupp05} and general reviews  \cite{Libro1,Libro2,Libro3}).

The feature of central syndrome that the improvement of perception by cross-modal effects
was greater as the primary stimulus to be perceived was weaker, is in
relation to what is observed later by other authors \cite{Poggel06,Laurienti06,Peiffer07}. Also, the fact this type of facilitation is greater as the cerebral lesion (excitability deficit) is greater, is in agreement with subsequent observations
\cite{Gillmeister07,Stein09,Schnupp05}. In a previous work \cite{Gonzalo03}, we suggested that the multisensory
interaction contributes nonlinearly with the stimuli to the cerebral excitation, which is
related to other suggestions \cite{Standford07}.

Concerning the facilitating effect of muscular effort in normal subjects, a slight improvement of
postural equilibrium by clenching fists was reported in Ref.
\cite{Ballus70}. Also, we have recently observed \cite{Neurocomputing13,Iwinac11} improvement of visual Vernier acuity (capability to discerning misalignment between two segments) by moderate static muscular effort (males holding 14 kg and females 9 kg) in 10 tested normal subjects (age range 21-61). This acuity is called also hyperacuity because  it is a higher function,  then it easily degradable according to what said in the previous
section.
The maximum improvement was found in the subject of older age  while for the youngest there was no conclusive improvement. This is in agreement
with the assumption of some deficit of cerebral excitability in the oldest subject
for which facilitating effects would be then more pronounced \cite{Laurienti06}.

The improvement of perception by facilitation can be considered a sensory ``growth" by means of an increase in the cerebral excitation which compensates in part for the neural mass lost in the central lesion, making the cerebral system to become more rapid and excitable \cite{Gonzalo10,Gonzalo51,Gonzalo52}.
This consideration is of special interest in relation to the following discussion on Stevens' power law.
First, we note that the validity of the Stevens's power law is assumed to be restricted to limited ranges of stimuli, and it is not
exempt from criticism. However, it is remarkable that many biological observable quantities are
statistically seen to scale with the {\em mass} $M$ of the organism, according to power laws $Y= k M^r$, where most of the
exponents $r$ are surprisingly found to be multiples of $1/4$ such as the metabolic rate ($r\approx 3/4$), lifespan ($r\approx 1/4$), growth rate ($r\approx -1/4$), height of trees ($r\approx 1/4$), cerebral gray matter ($r\approx 5/4$), etc.  These scaling laws are supposed to arise from universal mechanisms in all biological
systems as the optimization to regulate the activity of its subunits, as cells
(\cite{West05,Anderson01} and references therein). Under the assumption that a stimulus $S$
activates a neural mass $M_{neur}= \alpha S^{\beta}$
\cite{Arthurs04,Gonzalo07P}, the power law of perception
(\ref{STEVENS}) becomes $P= k (M_{neur})^{m/\beta}= k
(M_{neur})^r$ with $ r\equiv  m/\beta$, i.e., we recover a biological scaling power law of
growth which would be the basis of the Stevens' power law. Note that the exponent, and hence the
slope, is independent of the units used and other proportionality factors. In cases where $\beta$ would be close to unity
\cite{Arthurs04}, then $r\approx m$, and Stevens' law would exhibit quarter powers as seen in some of the cases analyzed here.
Thus, the sensory growth by means of intersensory or cross-modal facilitation would be a particular case of the universal biological growth laws.

\section{Improving Perception of a Stimulus by Increasing Its Intensity. More Power Laws}

As stated previously, the anomalies in central syndrome are clearly manifested under low stimulation, giving place to a decomposition or dissociation of the perception in which the  functions are lost according to allometric laws (see section 4). Now, let us consider the improvement of perception in central syndrome by increasing the intensity of the stimulus to be perceived. We use available data from the central syndrome cases, which are exceptional examples of quantification of perception \cite{Gonzalo10}, and again we use Stevens' power law given by Eq. (\ref{STEVENS}) in the fitting to data \cite{Gonzalo09,Iwinac09}.

\begin{figure}
\begin{center}
\hspace*{0.5cm}\includegraphics[width=6.5cm]{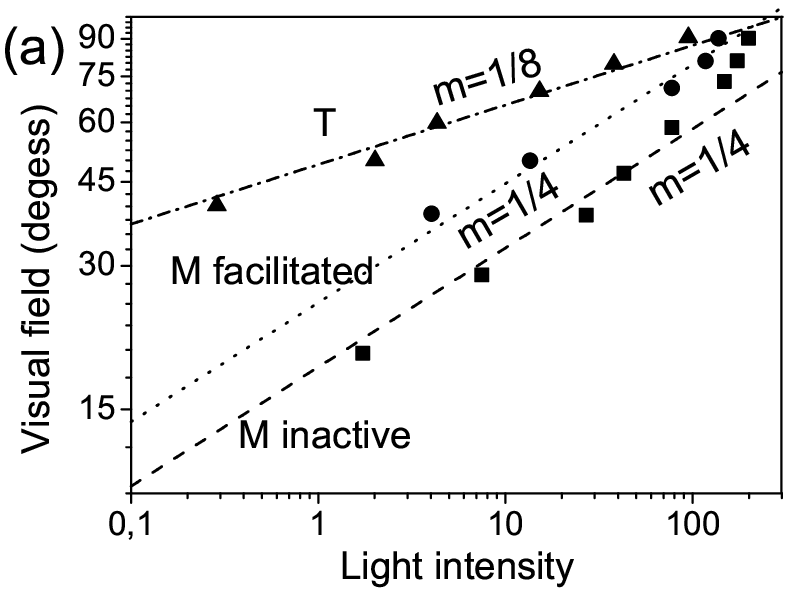}
\hspace*{0.3cm}
\includegraphics[width=6.5cm]{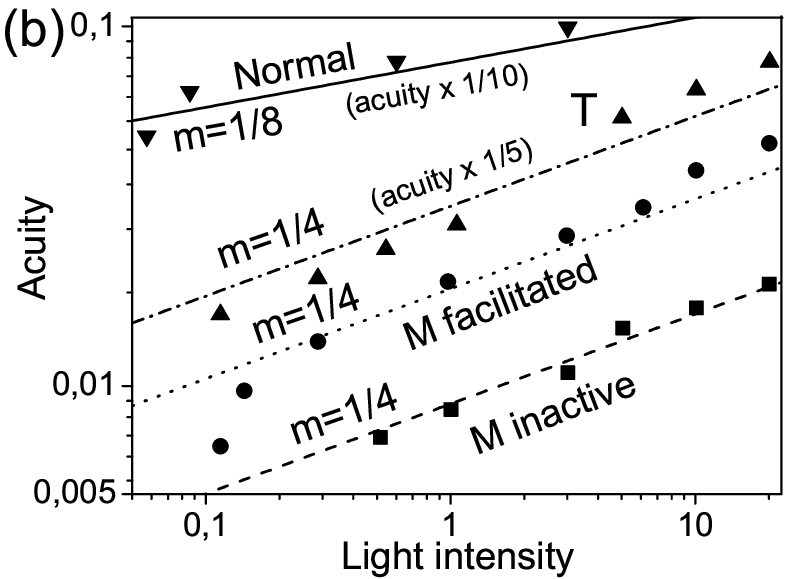}
\end{center}
\caption{\label{PLAWSTI} Log-log representation of: (a) Visual field amplitude of right eye versus
 illumination of test white disk of 1 cm diameter, for M inactive, M facilitated and  T.  (b) Acuity of right eye, direct vision, versus illumination, for M inactive, M facilitated, T and
for a normal subject.  (Adapted from the respective Fig. 8(a) and (b) of \cite{Gonzalo09} with permission of Elsevier). }
\end{figure}

 We show in Fig. \ref{PLAWSTI} (a), in log-log representation, the visual field amplitude of right eye versus
illumination intensity of a test white disk of 1 cm diameter. The exponents of the fitting are $m \approx 1/4$ for case M inactive (free of facilitation), $m \approx 1/4$ for M facilitated by constant muscular effort and $m \approx 1/8$ for the less acute case T. For visual acuity, we show in Fig. \ref{PLAWSTI} (b) the data and the fitting curves for direct vision of the right eye in case M inactive, M facilitated, T inactive and  a normal subject, versus light intensity of the stimulus.  The respective power exponents  are given.

However, the fitting corresponding to the lower data of visual acuity of case M facilitated is not good.
Contrary to the improvement of perception
versus facilitating stimulus, we found that improvement of perception
with the intensity of the primary stimulus is well-described in
some cases by two power laws, each one in the limits of very low and very
high stimulus intensity.
Although two branches can be also appreciated under other representations (Fechner Law for example) in such cases, we use scaling power laws because we consider they have more physical and biological meaning in relation to the dynamics of neural networks and to the laws of biological growth.

In order to account mathematically for two limiting power laws in a single expression we use a fitting function of the type
\begin{equation}\label{FUNC}
P = \left[1-\exp\left(-\frac{p_1}{p_2}S^{m_1-m_2}\right)\right]p_2S^{m_2}\, ,
\end{equation}
which reduces to the power law
\begin{equation}
P = p_1 S^{m_1}
\end{equation}
for very low intensities, and to the power law
\begin{equation}
P = p_2S^{m_2}
\end{equation}
for higher values of $S$. These two simple power laws are
represented by two straight lines in the graphs, where the exponents $m_1,\, m_2$ are indicated.

\begin{figure}
\begin{center}
\hspace*{0.5cm}\includegraphics[width=6.3cm]{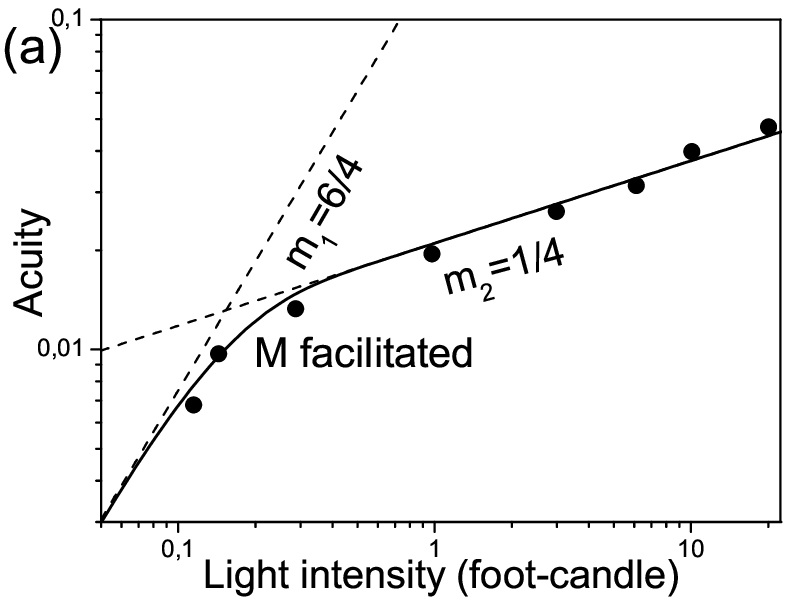}
\hspace*{0.3cm}
\includegraphics[width=6.5cm]{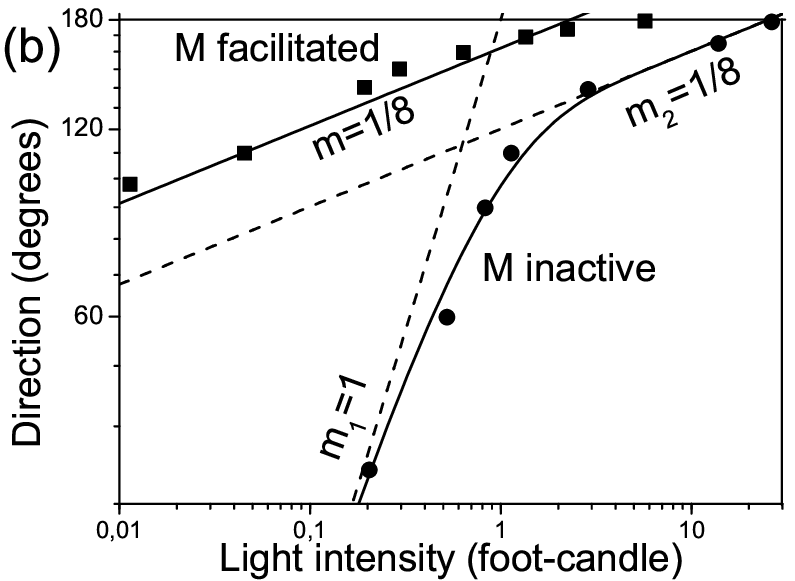}
\end{center}
\caption{\label{ACUITYGIRO} For direct vision of right eye of patient M,  log-log representation of:
(a) Acuity in facilitated state by a constant muscular effort, versus light intensity.
(b) Perceived direction of a vertical white 10 cm size test arrow, versus
light intensity illuminating the arrow on a black background, at 40 cm patient-arrow distance; for inactive state and facilitated state by muscular effort.}
\end{figure}

For the data of visual acuity of case M facilitated, a good fitting is obtained with the function given by Eq. (\ref{FUNC}), with exponent $m_1=6/4$ for low intensity and
$m_2=1/4$ for high intensity, as shown in Fig. \ref{ACUITYGIRO} (a).

The next example deals with the perception of a vertical upright white arrow by the right eye of the patient M in direct vision and for an inactive state.
The tilt was measured by rotating the arrow in the opposite direction until it was seen upright
($180^{\rm o}$). The direction perceived by M as a function of the light intensity that illuminated the arrow is shown in Fig.
\ref{ACUITYGIRO} (b). For the case of M inactive, the fitting with Eq. (\ref{FUNC}) gives $m_1=1$ for low light intensity and $m_2=1/8$
for high intensity. For M facilitated by muscular effort, a good fitting is possible with a single  power law with slope $m=1/8$.

\begin{figure}
\begin{center}
\includegraphics[width=6.5cm]{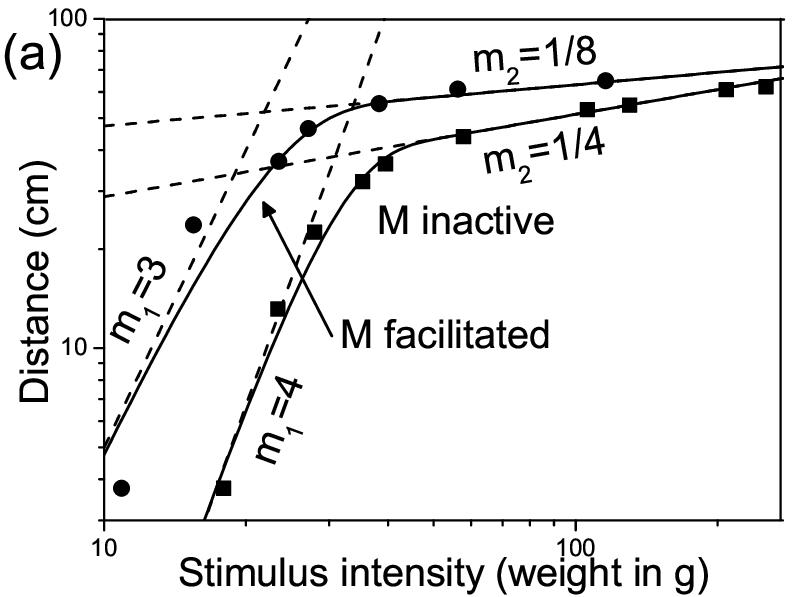}
\hspace{0.3cm}
\includegraphics[width=6.5cm]{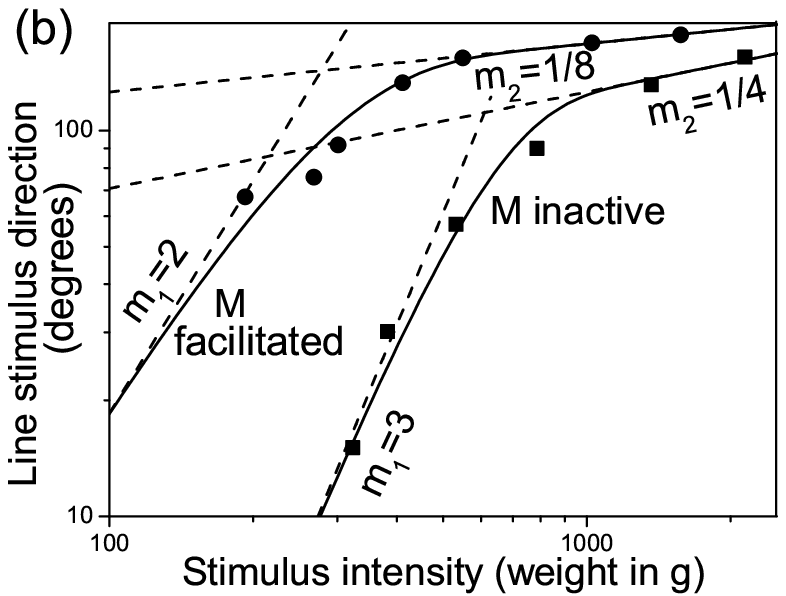}
\end{center}
\caption{\label{DISTANCEGIRO} In the log-log representation: (a) Distance in cm from the middle line
of the body to the localization of perception, as a function of the intensity (weight in $g$) of a punctual stimulus on the hand of
patient M, in inactive and facilitated states.
(b) Direction perceived of the line versus the pressure intensity (weight in $g$) on a line 6 cm long on one hand of patient M, in inactive and facilitated states.}
\end{figure}

In the tactile system, we first analyze the deviation of the perception of a punctual pressure stimulus on one hand towards the
middle line of the body. As shown in Fig. \ref{TACTILEINVER}, the perception in phase 2 is highly delocalized and centered at the
middle line of the body (completely reduced tactile field). The perception in phase 3 is contralateral and at a certain distance
from the middle line (a more developed tactile field). In phase 4, the perception becomes homolateral and still farther from the
middle line, until it reaches the normal localization on the hand in phase 5. Measurements of the deviation versus intensity of the
stimulus were made with the arms separated from the body and perpendicular to the middle line. The data and the fitting using a double-power-law
are shown in Fig. \ref{DISTANCEGIRO} (a) for M inactive and M facilitated. The exponents $m_1, m_2$ are indicated.

Now, we consider the case in which the  stimulus  is a pressure on a 6 cm long line on one hand. In this situation, there is also the
tilted perception of the orientation of the line stimulus, depending on the stimulus intensity (pressure of the line
stimulus measured in kg). We consider, as in inverted vision, that a tactile
direction of $180^{\rm o}$ means a correct (restored) perception of the line direction. The experimental data and the fittings with  Eq. (\ref{FUNC}) are shown
in Fig. \ref{DISTANCEGIRO} (b), where the exponents are indicated.

In general, it is noticeable that the exponents $1/4$ and $1/8$ appear equally in the examples for the tactile system (Fig. \ref{DISTANCEGIRO}) and for the visual system [Fig. \ref{PLAWSTI} (b) (c) and Fig. \ref{ACUITYGIRO}], for high enough stimulus.
An even more noticeable fact is that the exponents $1/4$ and $1/8$ also
appear in the power laws that relate the perception to the intensity of a facilitating stimulus (Fig. \ref{PLAWFACIL}).
This would suggest a similar neural network acting under high stimulus intensity and under a facilitating stimulus in a different modality. Thus, a higher stimulus would be
able to activate neurons with less specificity or with other specificities, in the same way as cross-modal facilitation does. Instead, a very weak stimulus (see left part
of Figs. \ref{ACUITYGIRO} and  \ref{DISTANCEGIRO}), would activate specific neurons mainly.

\section{Conclusion}

We have analyzed in a current context the central syndrome characterized and interpreted by Gonzalo \cite{Gonzalo10,Gonzalo51,Gonzalo52}. The first
remarkable feature of this syndrome is the multisensory involvement with symmetric bilaterality,
notwithstanding the unilateral lesion in a rather unspecific zone also referred to as the associative zone. The
proposed model of functional cortical gradients, which is an abstraction based on observed facts, accounts
for the great variety of different syndromes according to the position and magnitude of the lesion, and
explains the acute multisensoriality in the anomalies of the central syndrome. The model of gradients suggests
a functional continuity and unity of the cortex, as well as an integrative process through the whole gradient
extended over the cortex for a specific sensory function to be normal. Multisensory integration would be
involved in a greater degree in regions where the specific functional gradients overlap. This scheme responds to
requirements formulated recently, and to experimental findings in the last few years.

The central syndrome corresponds to a loss of neural mass in the region where the gradients overlap, producing a deficit of
cerebral excitability. In the new equilibrium reached, the syndrome is considered to be a scale reduction of the
cerebral system with respect to the normal one, exhibiting the same basic laws of nervous excitability but
on a smaller scale. This assumption permits a qualitative and quantitative interpretation of a second
relevant feature in the central syndrome: the dynamic decomposition or disgregation of perception into
its components as the nervous excitation diminishes (lower intensity of stimulus or greater lesion). Higher, more
complex functions, requiring greater nervous excitation, are lost before less complex functions, following a
well-defined physiological order. In this disgregation, the direction function can be discerned from others. It
can be lost in different degrees, giving place to the striking inverted perception disorder,
which we have related to other cases reported in the literature. The corresponding tactile and auditive inversion
disorders have not been described in the literature in the terms described in this research.

The progressive loss of the different qualities as the neural mass lost is greater would be governed by allometric scaling power laws. They are indeed seen to account quite accurately  for the continuous variation of a quality
(e.g., visual direction, acuity, etc.) with the variation of another quality (e.g., visual field). These continuous variations are governed by  power laws with  different power exponent for each quality. This quantification of the splitting of the perception
in different qualities has not yet been reported aside from this research, and stresses a functional unity and
continuity between the sensory functions (lower and higher) according to their excitability demands.

The functional unity of the cortex is also manifested in a third remarkable dynamic feature in the central syndrome: the capability to improve perception by means of another stimulus in the same or different (cross-modal) sensory modality. This secondary stimulus supplies an extra excitation that compensates in part the excitability deficit.
The permeability to cross-modal effects increases as the excitability deficit is higher, e.g., as the lesion
is greater, and as the primary stimulus is weaker. Similar statements can be found in more recent studies on the highly
active field of multisensory and cross-modal effects. This capability is then very low in normals, though
perceptible in some high functions under threshold stimulus. Facilitation by muscular effort, scarcely reported in the literature, is one of the most efficient ways to improve perception in the central syndrome. We have evidenced that moderate muscular
effort is also efficient in normals to improve Vernier acuity. The growth of the sensory level as a function of
the facilitating stimulus is found to follow approximately scaling power laws, which are supposed to emerge from
the dynamics of biological neural networks.

In the same way, the strong dependence of perception on the intensity of the primary stimulus to be perceived
is another pathway for sensory growth.  We have found that in some situations, perception versus stimulus can be described by two successive
power laws with different exponents. The exponent values for the range of high stimulus intensity are very similar for visual and tactile qualities, and also similar to those found for a facilitating stimulus. It would suggest
that for high intensity of
the stimulus, the involved neural network would be the same as for a facilitating
stimulus, i.e., the unspecific or multi-specific central zone.

To conclude, the central syndrome and its interpretation offer a framework that allows us to understand, at a functional level, several aspects of the distributed character of the cerebral processes. This framework is sustained by a physiological basis, related to the behavior of neural networks and  opens new avenues of research.

\end{document}